# EIR: Enhanced Image Representations for Medical Report Generation

Qiang Sun, Zongcheng Ji, Yinlong Xiao, Peng Chang and Jun Yu

*Abstract*—Generating medical reports from chest X-ray images is a critical and time-consuming task for radiologists, especially in emergencies. To alleviate the stress on radiologists and reduce the risk of misdiagnosis, numerous research efforts have been dedicated to automatic medical report generation in recent years. Most recent studies have developed methods that represent images by utilizing various medical metadata, such as the history of clinical documents with the current patient and the medical graphs constructed from the retrieved reports of other similar patients. However, all of the existing methods incorporate the additional metadata representations with visual representations through a simple "Add and LayerNorm" operation, which suffers from the information asymmetry problem due to the distinct distributions between them. In addition, the chest X-ray images are usually represented with pre-trained models based on images from natural domain, which exhibit an obvious domain gap between images from the general and medical domains. To this end, we propose a novel approach called enhanced image representations (EIR) for generating accurate chest X-ray reports. We utilize cross-modal transformers to combine the metadata representations with the image representations, thereby effectively addressing the information asymmetry problem between them, and we leverage pre-trained models from the medical domain to encode the medical images, effectively bridging the domain gap to represent the images. Experimental results conducted on widely used MIMIC and Open-I datasets show the effectiveness of our proposed method.

## I. INTRODUCTION

IN today's medical practice, especially during critical situations such as COVID-19 [1] or similar pandemics, a medical report serves as the primary medium for conveying the doctor's diagnosis [2]. Radiologists can analyze both normal and abnormal regions in radiology images from different views, utilizing their medical expertise and accumulated professional experience to write detailed reports [3]. However, this process is time-consuming and laborious for the radiologists. To alleviate the burden of increased demand for imaging examinations on radiologists and to assist less experienced radiologists in identifying abnormalities, there is a growing demand for research and development in the automatic generation of medical reports that demonstrate both the accuracy of clinical descriptions regarding the associated disease and corresponding symptoms, along with language fluency in generating realistic texts [4].

In recent years, fueled by the progress of image captioning [5], [6], which is a highly relevant task in computer vision, many approaches [7]–[11] have been proposed to generate medical reports automatically. These approaches follow principles similar to those employed in the image captioning task. The early methods, such as CNN-HRNN [12], [13], usually adopt an encoder-decoder structure to generate medical reports directly, where the image features are typically extracted by some typical Convolutional Neural Networks (CNNs) [14], [15] which conduct as the encoder and then fed into Recurrent Neural Networks (RNNs) [16] which conduct as the decoder to convert the visual features from the medical images to reports. Recently, to enhance the generation capabilities of the approach, there has been a shift in the choice of text decoder from RNNs to more powerful models, such as Long Short-Term Memory (LSTM) [17] and Transformers [18]. However, these image captioning methods [7]–[9] only consider images as input to generate simplistic descriptive sentences but disregard other available metadata, such as the history of clinical documents with the current patient and the existing reports of the other similar patients, which are crucial for producing comprehensive, contextually rich, and structured reports.

Recently, Nguyen *et al.* [19] proposed an approach to enhance the image representation aiming to generate more accurate reports by incorporating the historical medical reports of the current patient as additional input. Subsequently, Liu *et al.* [20] and Li *et al.* [21] utilized a universal graph constructed from the retrieved reports of other similar patients as additional input. However, all of these methods merely incorporate the representations of the additional metadata with visual representations through a simple "Add and LayerNorm" operation, which raises the issue of information asymmetry due to the distinct distributions between the image and other metadata representations. In addition, the effectiveness of

This work was supported by the Natural Science Foundation of China (62276242), National Aviation Science Foundation (2022Z071078001), CAAI-Huawei MindSpore Open Fund (CAAIXSJLJJ-2021-016B, CAAIXSJLJJ-2022-001A), Anhui Province Key Research and Development Program (202104a05020007), USTC-IAT Application Sci. & Tech. Achievement Cultivation Program (JL06521001Y), Sci. & Tech. Innovation Special Zone (20-163-14-LZ-001-004-01).(Corresponding author: Zongcheng Ji; Jun Yu)

Qiang Sun is with the Institute of Advanced Technology, University of Science and Technology of China, Anhui, Hefei 230027, China (e-mail: qiang1027@mail.ustc.edu.cn).

Zongcheng Ji and Peng Chang are with PAII Inc., California 94087, America (e-mail: jizongcheng@gmail.com; Pengchang@gmail.com).

Yinlong Xiao is with the Faculty of Information Technology, Beijing University of Technology, Beijing 100124, China (e-mail: yinlong.xiao@emails.bjut.edu.cn).

Jun Yu is with the Department of Automation and the Institute of Advanced Technology, University of Science and Technology of China, Anhui, Hefei 230027, China (e-mail: harryjun@ustc.edu.cn).

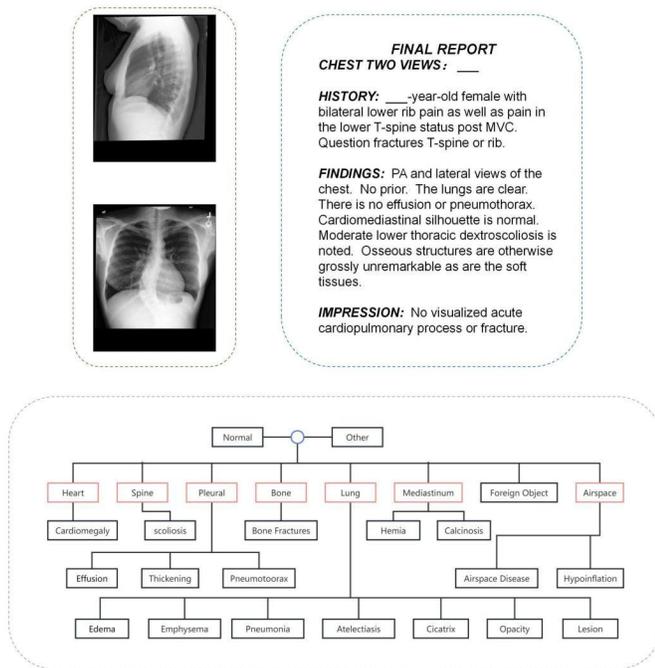

Fig. 1. An example of chest X-ray images and their metadata from MIMIC-CXR [24]. The metadata for images may include a medical report with multiple sections or a pre-constructed graph in [25]

incorporating all of different kinds of metadata as additional input for medical report generation remains unexplored in existing methods.

Furthermore, most existing methods [19], [22], [23] solely rely on the pre-trained models based on general domain images, such as ImageNet, to extract visual representations from images. However, they overlook the significant domain gap that exists between images from the general domain and the medical domain. As a result, these methods are unable to generate reports that accurately describe specific crucial abnormalities within the medical dataset. This constraint stems from the necessity for detailed recognition in medical tasks, the intricate and specialized nature of numerous complex medical terminologies, and the inadequate representation of medical images by pre-trained models trained on general domain data.

To tackle the mentioned concerns, we present an innovative approach called enhanced image representations (EIR) for generating accurate chest X-ray reports. Our approach integrates various metadata as supplementary input and leverages a pre-trained model specifically trained on medical images. Specifically, our approach consists of three main modules, i.e., the encoding module, the aggregation module, and the decoding module. The encoding module encodes the images, along with their metadata, which includes the clinical documents of the current patient and the universal graph constructed from the retrieved reports of other similar patients. We leverage pre-trained models that are specifically trained on medical domain data to encode the medical images, which enables us to effectively bridge the domain gap and ensure accurate representation of the medical images. The aggregation module utilizes cross-modal transformers to combine the metadata representations with the image representations, resulting in EIR. This approach effectively resolves the information asymmetry issue between the image and metadata representations, ensuring improved integration of different modalities. Finally, the decoding module employs a state-of-the-art classification-based report generation method [19] to generate fluent and more accurate medical reports using EIR.

In summary, our main contributions are as follows:

- We propose a novel approach called EIR to enhance the image representations for generating accurate chest X-ray reports, which integrates various metadata as supplementary input.
- We propose to utilize cross-modal transformers to combine the metadata representations with the image representations, thereby addressing the information asymmetry issue between them. Additionally, we propose to leverage pre-trained models from the medical domain to encode the medical images, effectively bridging the domain gap.
- We conduct extensive experiments on two widely used medical report generation datasets, demonstrating the effectiveness of our proposed method.

## II. RELATED WORK

### A. Image Captioning

In the field of natural image, image captioning is the most related computer vision task with the report generation task, both of which share the common goal of producing descriptive sentences derived from images. Most approaches [26], [27] used a traditional encoder-decoder structure to generate textual descriptions. But the framework could only generate concise sentences. Subsequently, there has been growing interest in research focusing on generating lengthy and semantically coherent paragraphs to describe images. J. Krause et al. [28] proposed a hierarchical recurrent network (HRNN), which comprises a paragraph RNN and a sentence RNN, and this two-step approach successfully addresses the limitation of generating longer text observed in previous text generation tasks. Recently Vaswani et al. [18] introduced a Transformer technique as an alternative to RNNs so that enhance the ability of the model to generate lengthy sentences and paragraphs. Parallel processing of sequential data is facilitated through the utilization of an attention mechanism, leading to the attainment of these outcomes. In contrast to Recurrent Neural Networks (RNN), transformers architecture process sentences holistically without recursion, allowing for a comprehensive exploration of word relationships beyond sequential constraints.

### B. Report Generation

Parallel to the success of image captioning related models ([29], [30], [31], [32], [33]), the research of medical report generation has also made impressive progress. Similar to image captioning task, most approaches have attempted to generate a fluent report automatically by adopting the chest X-ray images as input, excluding other metadata. In the pursuit of improving the clinical precision of the generated reports, various methods have been employed in generating complete medical reports. Li *et al.* introduced a hybrid model which are specialized in generating normal and abnormal sentences, respectively, thereby bolstering the model's capacity to describe abnormalities.. Jing *et al.* [34] adopted two RNNs to assist the whole model in generating more accurate sentences. At the same time, more researches have started to explore the capabilities of transformer. Marcella *et al.* introduced a mesh-connected transformer framework for image captioning in their work [35], utilizing encoder layer outputs effectively. Chen *et al.* [22] utilized a new model called memory-driven Transformer to generate radiology reports, incorporating relational memory and memory-driven conditional layer normalization to generate comprehensive reports with medical terms. While these approaches could generate fluent texts from images, these methods solely rely on images without additional metadata, limiting the expression of intended disease and symptom-related topics in reports which is a crucial aspect for physicians. Hence, it is crucial to investigate strategies for harnessing medical domain knowledge from diverse metadata to guide the process of report generation.

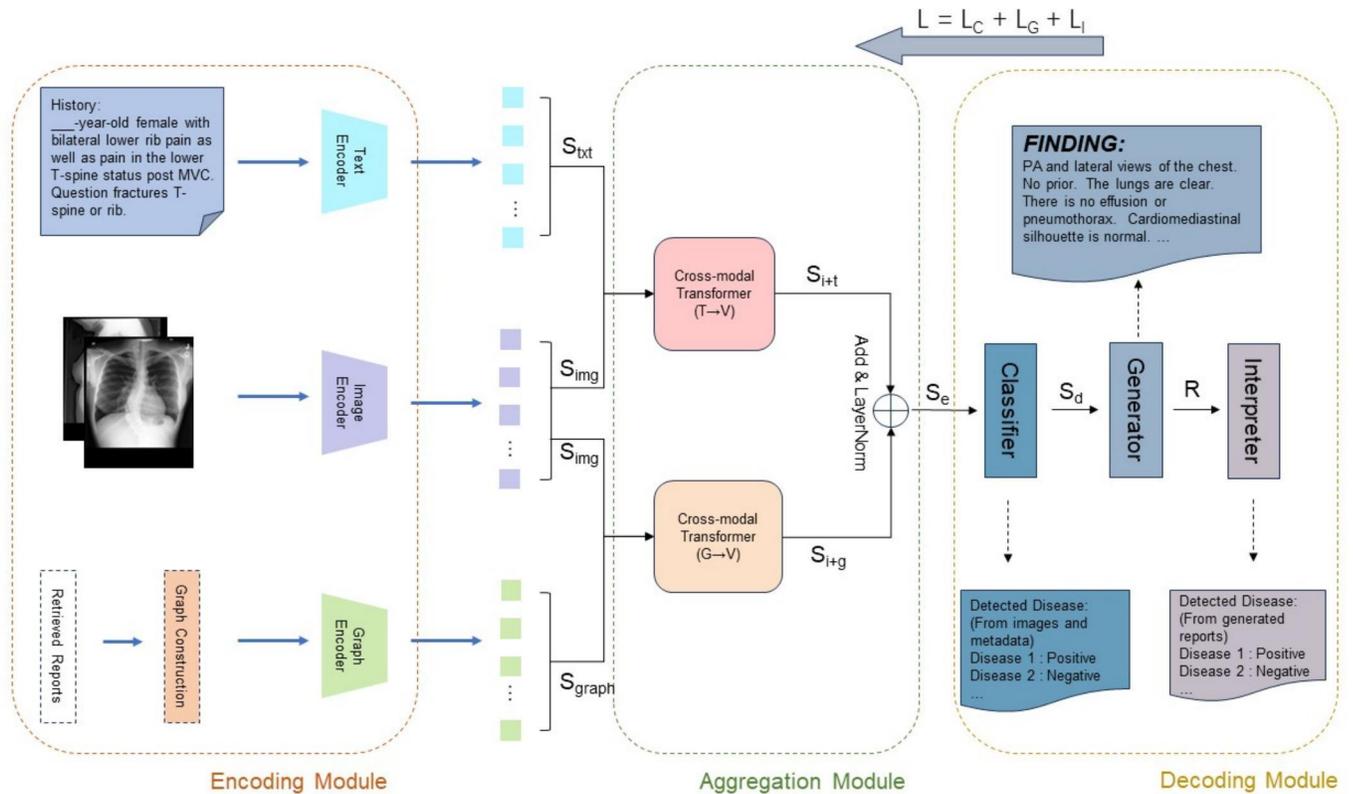

Fig. 2. Overview of our proposed approach, in which the encoding module explores the different metadata of the chest X-ray images, the aggregation module makes the representation have richer information by our Cross-modal Transformer and the decoding module generates a fluent and clinically accurate report with the framework from the state-of-the-art model [19].

### C. Report Generation Meets Metadata

As illustrated in Fig 1, a medical report typically comprises several sections, *e.g.*, history, impression and findings. Many studies have merged the impression and findings sections as the focal point for report generation tasks. This is done to mimic the working patterns of radiologists and utilize diverse metadata from the medical domain for report generation, the effectiveness of various kinds of metadata about the chest X-ray image has started to be explored and been utilized in the approach of medical report generation. Li *et al.* [36] implemented a graph-based methodology in which enhancing the generation process by transforming common abnormalities into medical report templates. This method also utilized a graph structure representing common abnormalities as prior knowledge for report generation. The subsequent researches [37], [38] have adopted and extended this graph structure to enhance the generation of report. Later Zhang *et al.* [25] implemented a universal graph containing 20 entities., which is pre-constructed using prior medical knowledge, to model relationships among entities linked to the same organ. Meanwhile, Liu *et al.* [20] proposed an approach to generate reports with more abnormalities by distilling the useful prior and posterior knowledge, i.e., the pre-constructed knowledge words and graphs. However, these methods utilize pre-constructed knowledge from metadata, which remains static during training, limiting adaptability for special cases. Therefore, there are also studies that integrate dynamic medical knowledge from metadata about the chest X-ray image to address this limitation. Nguyen *et al.* [19] proposed an end-to-end framework containing three complementary modules, which has achieved state-of-the art performance in generating fluent and clinically accurate radiology reports. This approach could learn the medical knowledge by making full use of metadata about the chest X-ray image as input for the model, including all images of different views and the history section from the medical report describing the patient's clinical history and indication from doctors. Li *et al.* [21] introduced an approach to enhance the generation of higher-quality reports by utilizing a dynamically updatable graph. This graph can be enriched with newly extracted knowledge obtained from the retrieved reports for each specific case.

To enhance visual feature representations by right means for the report generation and explore the effectiveness of incorporating both of these all kinds of metadata as additional input, our approach utilizes an encoding module to encode the different kinds of metadata about the chest X-ray image, *i.e.* the patient's clinical history from the rest sections of the medical report except for the most important sections that describe the corresponding image conditions and relevant symptoms and a universal graph metadata constructed from the retrieved reports of other similar patients. Through an aggregation module, our approach fuses the features from the different metadata about the chest X-ray images with visual features by our Cross-modal Transformer and then produces high-quality reports by a decoding module from the state-of-the-art model [19]. What's more, we propose to leverage pre-trained models from the medical domain to encode the medical images to further enhance visual feature representations.

## III. METHODS

Fig 2 illustrates the overview of our approach called EIR which follows an encoder-decoder architecture by enhancing the image feature representations to generate the high-quality medical report. The approach comprises three components, namely an encoding module, an aggregation module and a decoding module. The encoding module adopts different modality encoders to extract feature representations from different metadata, such as clinical documents from the reports and a universal graph constructed from the retrieved reports of other similar patients. With the goal of enhancing the image feature representations, we also extract image features with the pretrained model provided by a novel self-supervised pretraining approach from the medical domain [39]. All the feature representations are encoded into low-dimensional feature embeddings individually and then fed into the aggregation module as input. The aggregation module will repeatedly reinforce the visual features with the low-level feature representations from the different metadata about the chest X-ray images using Cross-modal Transformers, respectively. Finally, the end-to-end classifier-generator-interpreter framework from the state-of-the-art model introduced by Nguyen *et al.* [19] is employed as the decoding module to generate high-quality reports with the enhanced visual representations.

### A. Encoding Module

**Image Feature Extraction:** We denote the input chest X-ray images by $\{X_n\}_{n=1}^m$ where m is the number of images. In consideration of the obvious domain gap between natural images and medical images, We suggest employing a pre-trained ResNet-50 model for extracting image features. This model is initially trained using a momentum-based teacher-student architecture introduced by Zhou *et al.* [39]. The emphasis of the model lies in feature-level contrast, and it generates both homogeneous and heterogeneous data pairs by combining image and feature batches. Finally, we adopt the student network as the medical image encoder to get the medical image feature representation $\{x_n\}_{n=1}^m \in R^c$ from the images. Here c is the number of the features. Then, the multi-view image features $x \in R^c$ can be obtained by max pool layer across the set of m image features $\{x_n\}_{n=1}^m$, as proposed in Su *et al.* [40]. And then the extracted image feature representations $S_{img}$ will be fed to our enhancement module.

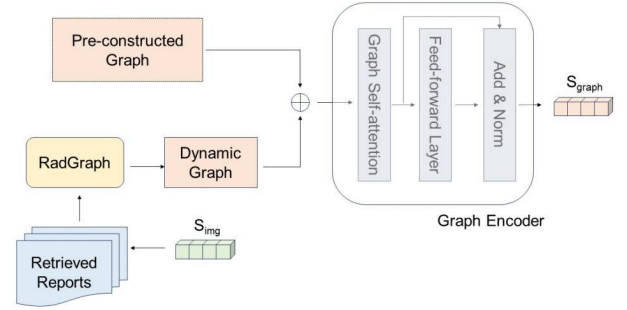

Fig. 3. Illustration of our process for graph feature extraction. The structure of the pre-constructed graph can be found in Fig 1

$$x_i = \text{ResNet50}(X_n) \\ S_{img} = \text{Maxpooling}(x_n) \quad (1)$$

**Text Feature Extraction:** As for metadata of textual modality, our approach employs the patient's clinical history from the rest sections of the medical report except for *FINDINGS* and *IMPRESSION* sections as text input. Let us denote a history text input as $T = \{w_1, w_2, ..., w_l\}$, where $w_i$ is the word embedding of the i-th word in the text. We utilize the Transformer encoder [18] consisting of the Multi-Head Attention (MHA) and Feed-Forward Network (FFN) to get the feature representation from the text metadata. And then a text document will be converted to $H = \{h_1, h_2, ..., h_l\}$ across the hidden layers in the encoder. Then the text feature embedding H that derived from the hidden layers will be integrated with text embedding $Q = \{q_1, q_2, ..., q_n\}$, which represents n disease-related topics to acquire the final text feature representation $S_{txt}$.

$$S_{txt} = \text{Softmax}(QH^T)H. \quad (2)$$

Here, $Q \in R^{n*e}$ is a randomly initialized matrix learned through an attention mechanism regarding relevant diseases. This way, The model can acquire a text feature representation that emphasizes the most pertinent words, such as *cough* or *shortness of breath*, which can enhance the model's generation of reports, particularly in terms of generating key medical terminology.

**Graph Feature Extraction:** Our approach also incorporates the dynamic chest knowledge graph proposed by Li *et al.* [21] as another type of metadata about the chest X-ray images for enhancing the visual feature representation from the chest X-ray images. Fig 3 illustrates the entire process of the construction and feature extraction for the dynamic graph metadata. First, we employ a bottom-up framework to build the fundamental structure of the graph metadata. Based on the input chest X-ray images, we retrieve the top-$n_T$ similar reports and update the graph structure with specific knowledge. This process yields the input structure for the knowledge graph metadata, comprising 28 entity nodes. The structure includes a global node representing a [CLS] token, 7 nodes representing organs or tissues, and the remaining 20 nodes representing diseases. Subsequently, we utilize the dynamic graph encoder in [21] as our graph encoder to extract knowledge graph feature representation. This encoder is built based on the Transformer encoder [18]. The process of graph feature extraction can be defined as follows:

$$\begin{aligned} \mathbf{S}_{graph} &= \text{LN}(\text{FFN}(\mathbf{e}_{gsa}) + \mathbf{e}_{gsa}) \\ \mathbf{e}_{gsa} &= \text{GSA}(\mathbf{f}_N, \mathbf{A}) + \mathbf{f}_N \end{aligned} \quad (3)$$

where GSA is a Multi-Head Attention module for the structural information of a knowledge graph. $\mathbf{f}_N$ is the initialized nodes representations and $\mathbf{A}$ is an adjacency matrix for the visible mask of the knowledge graph.

### B. Aggregation Module

The features extracted from metadata of different modalities often suffer from information asymmetry issues due to the simplistic "Add LayerNorm" operation.

To reinforce a *image modality representation* repeatedly with the low-level feature representations from another *metadata modality representation*, we propose to utilize an aggregation module with the Cross-modal Transformer model rather than via a simple "Add & LayerNorm" operation. As shown in Fig 4, the centerpiece of our Cross-modal Transformer model is the cross-modal attention module.

**Cross-Modal Attention:** We consider two modalities originating from distinct source metadata, denoted as α and β. For each of them, we also denote sequences of metadata represented by $\mathbf{S}_\alpha \in \mathbb{R}^{T_\alpha \times d_\alpha}$ and $\mathbf{S}_\beta \in \mathbb{R}^{T_\beta \times d_\beta}$, respectively. We propose to enhance one type of feature representation with another type of feature representation by providing a latent adaptation across modalities.

We define the way in which the cross-modal attention mechanism operates as follows:

$$\begin{aligned} S_{\alpha+\beta} &= \text{CA}_{\beta \to \alpha}(S_\alpha, S_\beta) \\ &= \text{softmax}\left(\frac{Q_\alpha K_\beta^\top}{\sqrt{d_k}}\right) V_\beta \\ &= \text{softmax}\left(\frac{S_\alpha W_{Q_\alpha} W_{K_\beta}^\top S_\beta^\top}{\sqrt{d_k}}\right) S_\beta W_{V_\beta}. \end{aligned} \quad (4)$$

where the Query, Key, Value pairs are defined as $\mathbf{S}_\alpha \mathbf{W}_{Q_\alpha}$, $\mathbf{S}_\beta \mathbf{W}_{K_\beta}$ and $\mathbf{S}_\beta \mathbf{W}_{V_\beta}$. The $\mathbf{S}_{\alpha+\beta}$ means the latent adaptation from β to α.

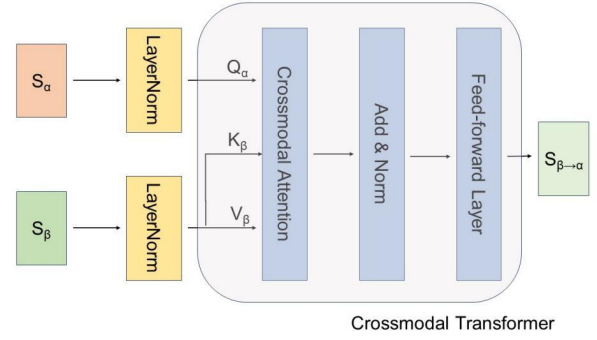

Fig. 4. Illustration of our proposed Cross-modal Transformer Model

**Cross-Modal Transformer:** In line with earlier research on transformers [18], [41], we introduce a residual connection into the computation of cross-modal attention. Subsequently, as illustrated in Figure 4, we incorporate an additional position-wise feed-forward sublayer to form a *Cross-modal Attention*. Each cross-modal attention module directly adjusts based on the low-level feature sequence.

Building upon the proposed cross-modal attention mechanism, we enhance the traditional Transformer to enable the features of the image modality to be strengthened by receiving valuable information from another metadata modality within the model. In the following sections, we illustrate an example of enhancing image feature representation with text feature representation, denoted as "T → V". Each cross-modal transformer is composed of D layers of cross-modal attention modules. Formally, the computation of the cross-modal transformer proceeds in a feed-forward manner for i = 1,..., D layers.:

$$\begin{aligned} S_{T \to V}^{[0]} &= S_V^{[0]} \\ \hat{S}_{T \to V}^{[i]} &= \text{CA}_{T \to V}^{[i], \text{mul}}(\text{LN}(S_{T \to V}^{[i-1]}), \text{LN}(S_T^{[0]})) \\ S_{T \to V}^{[i]} &= f_{\theta_T^{[i]} \to V}(\text{LN}(\hat{S}_{T \to V}^{[i]}) + \hat{S}_{T \to V}^{[i]}) \end{aligned} \quad (5)$$

where $f_\theta$ is a positionwise feed-forward sublayer parametrized by θ, and $\text{CA}_{T \to V}^{[i], \text{mul}}$ means a multi-head version of $\text{CA}_{T \to V}$ at layer i. LN means layer normalization.

In the aggregation module, the input feature representations of the image modality will be enriched through the cross-modal attention mechanism, incorporating information from other modalities to enhance the fine-grained representation of image features. In the proposed Cross-modal Transformer, metadata features from other modalities undergo transformation into a format suitable for interaction with image feature representations. We observed that the cross-modal transformer effectively learns correlations between feature representations from different modalities and enhances the feature information of the target modality. Empirically, our observations suggest that the cross-modal transformer adeptly learns to establish correlations among meaningful elements across modalities. The final Cross-modal Transformer (CT) is based on modeling every pair of cross-modal interactions. And then the enhanced image representation $S_{i+t}$ and $S_{i+g}$ are entangled to form *enriched disease representation* $S_e$ as

$$S_e = \text{LayerNorm}(S_{i+t} + S_{i+g}) \qquad (6)$$

Intuitively, through the proposed aggregation module, our approach enables the comprehensive utilization of all metadata information, simulating the workflow in a hospital. Radiologists generate accurate medical imaging reports based on patients' clinical history and historical experiences with similar cases. The proposed aggregation module, as indicated by the ablation study in the IV and V, has demonstrated a significant performance enhancement in the task of medical report generation.

### C. Decoding Module

For the decoding module in our approach, we use the framework from the state-of-the-art model proposed by Nguyen et al. [19] to generate medical report which mimics the hospital work flow by screening disease visual representations conditioned on clinical history or doctors' experience. The framework includes three complementary blocks: a classifier block generates the classification result embedding of disease-related topics by taking the enhanced image feature representation as input, and then a generator block generates the medical reports with the embedding representation from the classifier block. At the same time, an interpreter block could ensure the generated reports' consistency with respect to disease-related topics.

**Classifier Block:** The classifier block receives feature representations $S_e$ enhanced by the aggregation module as input and generates feature representation $S_d$ containing rich disease information for the accurate generation of subsequent text reports. Specifically, the classifier block will further encode information regarding the disease state after the fusion of features from different input metadata. This allows the model to obtain confidence scores for each disease and the classification loss is computed as following:

$$\begin{aligned} p &= \text{Softmax}(S_e S^\top), \\ \mathcal{L}_\mathcal{C} &= -\frac{1}{n} \sum_{i=1}^{n} \sum_{j=1}^{k} y_{ij} \log(p_{ij}). \end{aligned} \qquad (7)$$

where $S$ is randomly initialized and learned with above classification loss during training process. Finally, the *final disease embedding* $S_d \in \mathbb{R}^{n \times e}$ is the composition of disease state embedding $S_{states}$ about the status information of the disease and disease name embedding $S_{topics}$ about the thematic information of the disease which are all generated by the classifier block, as well as the disease representation embedding $S_e$ about the characteristics information of the disease that are output from the aggregation module.

$$S_d = S_{states} + S_{topics} + S_e. \qquad (8)$$

In this way, the learned and obtained disease embedding contains rich disease details information, closely aligning with the process of radiologists writing diagnostic reports in reality. This aims to enhance the accuracy of automatically generated text reports.

TABLE I
THE STATISTICS OF THE TWO BENCHMARK DATASETS

| DATASET | Open-I | | | MIMIC-CXR | | |
|---|---|---|---|---|---|---|
| | TRAIN | VAL | TEST | TRAIN | VAL | TEST |
| IMAGE | 5226 | 748 | 1496 | 368960 | 2991 | 5159 |
| REPORT | 2770 | 395 | 790 | 222758 | 1808 | 3269 |
| PATIENT | 2770 | 395 | 790 | 64586 | 500 | 293 |

**Generator Block:** Later the generator block that derived from the transformer decoder produces the medical reports by taking the final enriched disease embedding representation $S_d$ as input. Subsequently, the model predicts each word in the final text report based on the hidden layers within its own architecture. We define $p_{\text{word},ij}$ as the confidence value for selecting the j-th word from the vocabulary at the i-th position in the generated text report. The loss from the generator block is calculated as the cross-entropy between the words $y_{\text{word}}$ from the ground-truth reports and the words $p_{\text{word}}$ from the model:

$$\mathcal{L}_\mathcal{G} = -\frac{1}{l} \sum_{i=1}^{l} \sum_{j=1}^{v} y_{\text{word},ij} \log(p_{\text{word},ij}). \qquad (9)$$

Later, based on the confidence values, the corresponding embedding for the generated report can be obtained for further fine-tuning by the subsequent interpretation block.

**Interpreter Block:** In this model architecture, the generated medical reports may still exhibit inaccuracies or may not align with the disease theme. This manifests as a mismatch between the generated medical reports and the results from the classification block. Therefore, an interpreter block is incorporated into the overall architecture. This interpreter block classifies based on the output of the generator block, providing a feedback loop to adjust the generated medical reports by comparing them with the output from the classification block. Throughout the entire training process, the parameters of the interpreter block remain frozen. When the medical theme associated with the generated medical report is inconsistent with the initial medical theme corresponding to the classification block, the interpreter block employs the following multi-label classification loss $L_I$. This compels the model to generate the final medical report in a direction consistent with the medical theme. Ultimately, the model from the state-of-the-art model proposed by Nguyen et al. [19] generates the final text report based on the total loss $L_{total}$.

$$\mathcal{L}_\mathcal{I} = -\frac{1}{n} \sum_{i=1}^{n} \sum_{j=1}^{k} y_{ij} \log(p_{\text{int},ij}). \qquad (10)$$

$$L_{total} = L_C + L_G + L_I.$$

here $y_{ij}$ and $p_{\text{int},ij}$ individually represent the disease label from ground truth and our interpreter block.

### IV. RESULTS AND DISCUSSION

To illustrate the effectiveness of our proposed approach, this section provides qualitative analysis and quantitative analysis of our experiment results on two public datasets, MIMIC-CXR [24] and Open-I [42].

TABLE II
THE PERFORMANCCE OF OUR PROPOSED APPROACH AND MANY EXISTING METHODS ON OPEN-I AND MIMIC-CXR DATASET. THE BEST RESULTS ARE HIGHLIGHTED IN **BOLD FACE**. DIFFERENT LANGUAGE METRICS ARE EMPLOYED: BLEU-1 TO BLEU-4 (BL-1 TO BL-4), AND ROUGE-L (RG-L)

| Datasets | Methods | NLG Metrics | | | | |
|---|---|---|---|---|---|---|
| | | BL-1 | BL-2 | BL-3 | BL-4 | RG-L |
| Open-I | S&T [5] | 0.316 | 0.211 | 0.140 | 0.095 | 0.267 |
| | SA&T [30] | 0.399 | 0.251 | 0.168 | 0.118 | 0.323 |
| | R2Gen [22] | 0.470 | 0.304 | 0.219 | 0.165 | 0.371 |
| | KERP [36] | 0.482 | 0.325 | 0.226 | 0.162 | 0.339 |
| | HRGR [11] | 0.438 | 0.298 | 0.208 | 0.151 | 0.322 |
| | MKG [25] | 0.441 | 0.291 | 0.203 | 0.147 | 0.367 |
| | CA [43] | 0.492 | 0.314 | 0.222 | 0.169 | 0.381 |
| | PPKED [20] | 0.483 | 0.315 | 0.224 | 0.168 | 0.376 |
| | MGSK [44] | 0.496 | 0.327 | 0.238 | 0.178 | 0.381 |
| | Nguyen [19] | 0.515 | 0.378 | 0.293 | 0.235 | 0.436 |
| | EIR | **0.521** | **0.388** | **0.306** | **0.249** | **0.464** |
| MIMIC-CXR | SA&T [30] | 0.370 | 0.240 | 0.170 | 0.128 | 0.310 |
| | AdpAtt [45] | 0.384 | 0.251 | 0.178 | 0.134 | 0.314 |
| | R2Gen [22] | 0.353 | 0.218 | 0.145 | 0.103 | 0.284 |
| | CMN [46] | 0.353 | 0.218 | 0.148 | 0.106 | 0.278 |
| | M2TR [23] | 0.378 | 0.232 | 0.154 | 0.107 | 0.272 |
| | GumbelTrans [10] | 0.415 | 0.272 | 0.193 | 0.146 | 0.318 |
| | CA [43] | 0.350 | 0.219 | 0.152 | 0.109 | 0.283 |
| | PPKED [20] | 0.360 | 0.224 | 0.149 | 0.106 | 0.284 |
| | MGSK [44] | 0.363 | 0.228 | 0.156 | 0.115 | 0.284 |
| | Nguyen [19] | 0.495 | 0.360 | 0.278 | 0.224 | 0.390 |
| | EIR | **0.515** | **0.382** | **0.310** | **0.248** | **0.415** |

TABLE III
QUANTITATIVE COMPARISON OF CLINICAL ACCURACY ON MIMIC-CXR DATASET.

| Methods | Precision | Recall | F1-score |
|---|---|---|---|
| TopDown [26] | 0.166 | 0.121 | 0.133 |
| R2Gen [22] | 0.333 | 0.273 | 0.276 |
| MGSK [44] | 0.458 | 0.348 | 0.371 |
| Nguyen [19] | 0.448 | 0.399 | 0.407 |
| EIR | 0.450 | 0.437 | 0.443 |

### A. Datasets

As shown in Table I, we validate our proposed method on the two widely used benchmark datasets, following the settings described in [22] to partition the datasets, ensuring fairness in the comparison

(1) The Open-I Dataset, a public radiology dataset originated from the Indiana University hospital network with 3955 medical reports and 7470 images taken from different angles. The data for each patient may include multiple X-Ray images from different views and a medical report with multiple sections including *indication*, *comparison*, *findings* and *impression*. Here our approach utilizes the multi-view images as visual input and the text from the *indication* section as contextual metadata input. As for graph metadata input, we extract 739 unique entities from the cases with at least two images as dynamic node candidates by Stanta [47]. We will update the graph metadata when we train our model.

MIMIC-CXR is another large publicly available dataset used for medical report generation tasks. The specific format of this dataset is similar to the Open-I dataset, but it is larger in scale. The dataset comprises 473,057 chest X-ray images from different perspectives, labeled with classifications, and 236,563 medical reports from 73,478 patients.

(2) The MIMIC-CXR Dataset is another large publicly available radiology dataset used for medical report generation task, with 236563 medical reports of 73478 patients and 473057 images from multiple views, in a data format similar to the Open-I dataset. Each study for each patient may comprise chest X-ray images from multiple views and medical reports with following sections: *comparison*, *clinical history*, *indication*, *reasons for examination*, *impressions* and *findings*. In our approach, we adopt the multi-view images as visual input and the contextual information the concatenation of the *clinical history*, *reason for examination* and *indication* sections as contextual metadata input. But for graph metadata input, we get structural radiology information by RadGraph [48], which consists of 2895725 (*suggestive of*), 6115264 (*located at*) and 4010875 (*modify*) triplets classified by relation. We will also update the graph metadata during our training process.

TABLE IV
QUANTITATIVE ANALYSIS OF OUR APPROACH ON OPEN-I DATASET.

| Methods | NLG Metrics | | | | |
|---|---|---|---|---|---|
| | BL-1 | BL-2 | BL-3 | BL-4 | RG-L |
| M0 (MV) [19] | 0.476 | 0.324 | 0.228 | 0.164 | 0.379 |
| M1 (MV+T) [19] | 0.485 | 0.355 | 0.273 | 0.217 | 0.422 |
| M2 (MV+T+I) [19] | 0.515 | 0.378 | 0.293 | 0.235 | 0.436 |
| M3 (MV+T+CT) | 0.488 | 0.359 | 0.280 | 0.224 | 0.428 |
| M4 (MV+G) | 0.482 | 0.335 | 0.243 | 0.180 | 0.420 |
| M5 (MV+G+CT) | 0.484 | 0.349 | 0.251 | 0.180 | 0.426 |
| M6 (MV+T+G) | 0.506 | 0.375 | 0.289 | 0.229 | 0.438 |
| M7 (MV+T+G+CT) | 0.509 | 0.377 | 0.290 | 0.233 | 0.440 |
| M8 (MV+T+G+CT+DPM) | 0.515 | 0.381 | 0.294 | 0.237 | 0.442 |
| M9 (MV+T+I+G+CT+DPM) | 0.521 | 0.388 | 0.306 | 0.249 | 0.464 |

TABLE V
QUANTITATIVE ANALYSIS OF OUR APPROACH ON MIMIC-CXR DATASET.

| Methods | NLG Metrics | | | | |
|---|---|---|---|---|---|
| | BL-1 | BL-2 | BL-3 | BL-4 | RG-L |
| M0 (MV) [19] | 0.451 | 0.292 | 0.201 | 0.144 | 0.320 |
| M1 (MV+T) [19] | 0.491 | 0.357 | 0.276 | 0.223 | 0.389 |
| M2 (MV+T+I) [19] | 0.495 | 0.360 | 0.278 | 0.224 | 0.390 |
| M3 (MV+T+CT) | 0.493 | 0.360 | 0.279 | 0.225 | 0.393 |
| M4 (MV+G) | 0.485 | 0.332 | 0.245 | 0.189 | 0.375 |
| M5 (MV+G+CT) | 0.486 | 0.346 | 0.253 | 0.193 | 0.377 |
| M6 (MV+T+G) | 0.495 | 0.362 | 0.280 | 0.228 | 0.393 |
| M7 (MV+T+G+CT) | 0.505 | 0.369 | 0.288 | 0.230 | 0.395 |
| M8 (MV+T+G+CT+DPM) | 0.510 | 0.378 | 0.295 | 0.242 | 0.405 |
| M9 (MV+T+I+G+CT+DPM) | 0.512 | 0.381 | 0.301 | 0.245 | 0.410 |

### B. Evaluation Metrics

In our experiment, we assess the performance of our classification results using clinical efficacy (CE) metrics. These metrics include precision, recall, and F1 score. We use these metrics to compare the model-generated labels for 14 different categories related to thoracic diseases with the corresponding labels previously annotated by CheXpert [49] based on the reports. Notably, since the Open dataset was not labeled using CheXpert previously, we only report the model's results on the MIMIC dataset.

Simultaneously, we evaluate the overall performance of our approach using traditional natural language generation (NLG) metrics, specifically BLEU [50] and ROUGE-L [51]. BLEU, a primary NLG metric for assessing the quality of generated medical reports, was initially proposed for machine translation tasks. The evaluation involves assessing the word n-gram overlap between predictions and ground truth reports, measured using BLEU. Additionally, we employ ROUGE-L for comparison. Our approach exhibits noteworthy BLEU and ROUGE-L values.

### C. Main Results

*1) Language Generation Performance:* As shown in Table II, to demonstrate the effectiveness of our approach, we conduct experiments based on the most widely used language evaluation metrics and compared with many existing baseline models on the two benchmarks. S&T [5], SA&T [30] and AdpAtt [45] are three baselines used for image captioning task. R2Gen [25], KERP [36], HRGR [11], MKG [25], CA [43], CMN [46], M2TR [23], GumbelTrans [10] are the baselines used for medical report generation task that only adopt chest X-ray images as input. PPKED [20], Nguyen [19] and MGSK [44] is the baselines that take chest X-ray images and different metadata as input by a simple operation to enhance the input feature representations. All of the methods follow the same settings, we just cite the results from original papers.

The results in Table II show that our model outperforms the baselines on all of the language metrics. Higher BLEU and ROUGE-L values show that our approach could not only highlight key disease anomalies which is the most important for a successful medical report, but also reduce repetition of frequent sentences. With more metadata about the chest X-ray images as input and enhancing the visual representation by our Crossmodal Transformer, our approach has been far superior to the best SOTA methods of MGSK [44] on Open-I dataset and GumbelTrans [10] on MIMIC-CXR dataset.

*2) Clinical Accuracy Performance:* As depicted in Table III, we utilized the open-source CheXpert labeler tool [10] to assess the accuracy of our approach across 14 common diseases, evaluating the clinical precision of the generated reports. We compared F-1, Precision, and Recall scores of our EIR with those of other baseline models on the MIMIC-CXR dataset. Notably, our method exhibits a significant improvement in scores when compared to other existing methods. The results highlight that our approach achieves the highest Recall score and F1 score. In contrast to the state-of-the-art method MGSK [44], which leverages general prior knowledge and specific information from RadGraph [48], our approach demonstrates superior Recall and F1 scores. This improvement can be attributed to the incorporation of additional metadata related to chest X-ray images as input and the fusion of diverse feature representations facilitated by our aggregation module. What's more, the domain pretrained model used in the extraction of image feature also plays a crucial role.

### D. Ablation Study

To showcase the efficacy of each module in our proposed method, we have developed the following variations:

*M0-MV*: This primitive model adopts the multi-view images as input with the model from the approach proposed by Nguyen *et al.* [19].

*M1-MV+T*: In this model, contextual metadata is used with the model M0.

*M2-MV+T+I*: In this model, the interpreter block is used with the model M1 to improve the clinical accuracy by finetuning the generated report.

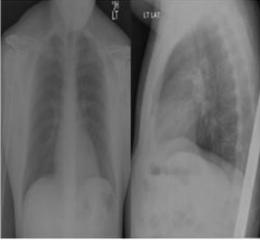
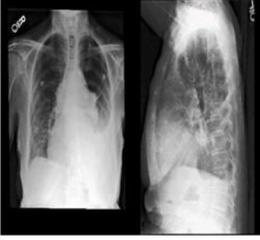

Fig. 5.  Examples of the generated reports on Open-I dataset and MIMIC-CXR dataset. We show the reports generated by EIR and its variants. For better visualization, We highlight the key words with different colors

*M3-MV+T+CT*: In this model, the Crossmodal Transformer is used with the model M1

*M4-MV+G*: In this model, graph metadata is used with the model M0.

*M5-MV+G+CT*: In this model, the Crossmodal Transformer is used with the model M4.

*M6-MV+T+G*: In this model, contextual metadata is used with the model M4.

*M7-MV+T+G+CT*: In this model, the Crossmodal Transformer is used with the model M6.

*M8-MV+T+G+CT+DPM*: In this model, the domain image pretrained model is used with the model M7.

*M9-MV+T+I+G+CT+DPM*: In this model, the interpreter block is used with the model M8.

Table IV and Table V have reported the quantitative analysis on Open-I dataset and MIMIC-CXR dataset, respectively. Our baseline is employed the framework proposed by [19].

**Effect of Encoding Module**. Different from the previous report generation methods, our approach employs the dynamic graph as a new metadata about chest X-ray images to enhance our input representation, what is more, we get our image representation with the pre-trained model based on medical demain. Our graph metadata could utilize the general knowledge and the specific knowledge. And the pre-trained model in the image feature extraction could learn the representation of medical images to make the final report incorporate the disease-related topics. As shown in Table IV and Table V, comparing *M0-MV+T* with *M4-MV+G* and *M6-MV+T+G*, we observe that incorporating more metadata as input effectively enhances the performance of the approach. Specifically, adopting dynamic graph as metadata improves the performance across all generation metrics. Additionally, comparing *M7-MV+T+G+CT* with *M8-MV+T+G+CT+DPM*, the results show that utilizing a pre-trained model specifically designed for the medical domain to extract image features also contributes to the enhancement of performance metrics. This outcome validates the effectiveness and necessity of the encoding module in the approach. The research results suggest that the performance improvement may be attributed to the ability to emphasize keywords in the generated reports, making them more closely aligned with ground truth reports.

**Effect of Aggregation Module**. As illustrated in Table IV and Table V, by comparing *M1-MV+T* and *M6-MV+T+G* with *M3-MV+T+CT* and *M7-MV+T+G+CT*, we evaluate the effectiveness of the different feature fusion methods. The results show that our proposed aggregation module can improve the language scores again. This we mainly attribute to the efficient enhancement for image representation. Previous studies just simply added representations of other metadata with image representation, which still remains the challenge of information asymmetry. However, the enhancement module we utilize mimics how radiologists write reports from the medical images and the metadata, which is crucial to improve the generated reports' quality. Hence, the generated reports receive higher language scores than other methods.

### E. Case Study

To delve deeper into the efficacy of our EIR, we perform a qualitative analysis on Open-I [42] dataset and MIMIC-CXR [24] dataset. Fig 5 exhibit the generated reports on the two datasets with all utilized inputs, including multi-view chest images, patients' clinical histories, and specific knowledge retrieved from similar reports, the reports from the ground truth, our model, and corresponding variants M3 and M5. Key medical terms in the reports were highlighted in different colors. The reports in two datasets contain frequently occurring medical terms. For these statements, all models can generate accurate descriptions. However, the effectiveness of our approach is observable, as some crucial medical terms are often present in various adopted metadata. Through our aggregation module, the model is able to focus on these key pieces of information, emphasizing them in the generated reports. It is why our approach EIR can predict sentences 'The visualized osseous structures of the thorax are without acute osseous abnormality'. The results show that our approach could generate the sentences that are unusual with more dynamic metadata as input and the Cross-modal Transformer.

## V. CONCLUSION

This paper introduces a novel approach named EIR, designed to enhance the representation of image features for the generation of more practical medical reports. The method achieves this goal by integrating diverse metadata, obtaining richer feature representations using models from the medical domain, and utilizing a cross-modal transformer to combine metadata representations with image representations. Experimental results indicate that, compared to other direct report generation methods, our approach significantly improves the description of crucial medical terms in the generated reports. Throughout the report generation process, our method not only extracts more comprehensive image feature representations for key terms but also adeptly integrates multi-modal features to produce medical reports that are clinically more accurate and linguistically more readable. Therefore, we illustrate that the model can generate high-quality medical reports, albeit recognizing there is still room for improvement.